\documentclass[preprint,journal]{vgtc}         

\usepackage{microtype}                 
\PassOptionsToPackage{warn}{textcomp}  
\usepackage{textcomp}                  
\usepackage{mathptmx}                  
\usepackage{times}                     
\usepackage{cite}                      
\usepackage{tabu}                      
\usepackage{booktabs}                  

\usepackage{listings}
\usepackage{supertabular}
\usepackage{enumitem}
\usepackage{multirow}
\usepackage{tabularx}
\usepackage{xcolor}
\usepackage{xspace}
\usepackage{amsmath}
\usepackage{color}
\usepackage{soul}
\usepackage{subfig}
\usepackage{caption}
\usepackage{graphicx}

\colorlet{punct}{red!60!black}
\definecolor{background}{HTML}{EEEEEE}
\definecolor{delim}{RGB}{20,105,176}
\colorlet{numb}{magenta!60!black}

\lstdefinelanguage{json}{
    basicstyle=\normalfont\ttfamily,
    numbers=left,
    numberstyle=\scriptsize,
    stepnumber=1,
    numbersep=8pt,
    showstringspaces=false,
    breaklines=true,
    frame=lines,
    backgroundcolor=\color{background},
    literate=
     *{0}{{{\color{numb}0}}}{1}
      {1}{{{\color{numb}1}}}{1}
      {2}{{{\color{numb}2}}}{1}
      {3}{{{\color{numb}3}}}{1}
      {4}{{{\color{numb}4}}}{1}
      {5}{{{\color{numb}5}}}{1}
      {6}{{{\color{numb}6}}}{1}
      {7}{{{\color{numb}7}}}{1}
      {8}{{{\color{numb}8}}}{1}
      {9}{{{\color{numb}9}}}{1}
      {:}{{{\color{punct}{:}}}}{1}
      {,}{{{\color{punct}{,}}}}{1}
      {\{}{{{\color{delim}{\{}}}}{1}
      {\}}{{{\color{delim}{\}}}}}{1}
      {[}{{{\color{delim}{[}}}}{1}
      {]}{{{\color{delim}{]}}}}{1},
}

\usepackage{fontawesome}

\newcommand{\remove}[1]{}

\newcommand{\revise}[1]{#1}

\newcommand{\attr}[1]{\emph{#1}}

\newcommand{\pc}[1]{$P_{C}$#1}
\newcommand{\pa}[1]{$P_{A}$#1}

\newcommand{\bpstart}[1]{
\noindent{\textbf{#1}}
}

\newcommand{\app}{Lumos\xspace}
\newcommand{\usera}{Austin\xspace}
\newcommand{\userb}{Taylor\xspace}
\newcommand{\userc}{Chandler\xspace}

\definecolor{orange}{RGB}{237,125,49}
\definecolor{purple}{RGB}{112,47,160}
\definecolor{pink}{RGB}{245,114,255}
\definecolor{black}{RGB}{0,0,0}

\ifpdf
  \pdfoutput=1\relax                   
  \usepackage{graphicx}                
  \DeclareGraphicsExtensions{.pdf,.png,.jpg,.jpeg} 
\else
  \usepackage{graphicx}                
  \DeclareGraphicsExtensions{.eps}     
\fi%

\graphicspath{{src/figures/}{pictures/}{images/}{./}} 

\ieeedoi{10.1109/TVCG.2021.3114827}

\usepackage{mathptmx}
\usepackage{graphicx}
\usepackage{times}
\usepackage{comment}

\usepackage[bookmarks,backref=true,linkcolor=black]{hyperref} 
\hypersetup{
  pdfauthor = {},
  pdftitle = {},
  pdfsubject = {},
  pdfkeywords = {},
  colorlinks=true,
  linkcolor= black,
  citecolor= black,
  pageanchor=true,
  urlcolor = black,
  plainpages = false,
  linktocpage
}

\onlineid{0}

\vgtccategory{Research}
\vgtcpapertype{system}

\title{\app: Increasing Awareness of Analytic \revise{Behavior} during Visual Data Analysis}

\author{Arpit Narechania, Adam Coscia, Emily Wall, Alex Endert}
\authorfooter{
\item
 Arpit Narechania, Adam Coscia, Alex Endert are with Georgia Institute of Technology. Emails: \{arpitnarechania, acoscia6, endert\}@gatech.edu. Emily Wall is with Emory University. Email: emily.wall@emory.edu *work performed while at Northwestern University.
}

\shortauthortitle{Narechania \MakeLowercase{\textit{et al.}}:\app: Increasing Awareness of Analytic Behavior during Visual Data Analysis}

\abstract{Visual data analysis tools provide people with the agency and flexibility to explore data using a variety of interactive functionalities. However, this flexibility may introduce potential consequences in situations where users unknowingly \remove{exhibit biases towards or against}\revise{overemphasize or underemphasize} specific subsets of the data or attribute space they are analyzing. For example, users may overemphasize specific attributes and/or their values (e.g., \revise{Gender is always encoded on the X axis}\remove{with the color encoding channel}), underemphasize others (e.g., Religion is never encoded), ignore a subset of the data (e.g., older people are filtered out), etc. In response, we present Lumos, a visual data analysis tool that captures and shows the interaction history with data to increase awareness of such analytic behaviors. Using in-situ (at the place of interaction) and ex-situ (in an external view) visualization techniques, Lumos provides real-time feedback to users \revise{for them to reflect on their activities}. For example, Lumos highlights datapoints that have been previously examined in the same visualization (in-situ) and also overlays them on the underlying data distribution (i.e., baseline distribution) in a separate visualization (ex-situ). 
Through a user study with 24 participants, we investigate how Lumos helps users' data exploration and decision-making processes. We found that Lumos increases users' awareness of visual data analysis practices in real-time, promoting reflection upon and acknowledgement of their intentions and potentially influencing subsequent interactions.
}


\CCScatlist{ 
 \CCScat{K.6.1}{Management of Computing and Information Systems}%
{Project and People Management}{Life Cycle};
 \CCScat{K.7.m}{The Computing Profession}{Miscellaneous}{Ethics}
}

\teaser{
  \includegraphics[width=\linewidth]{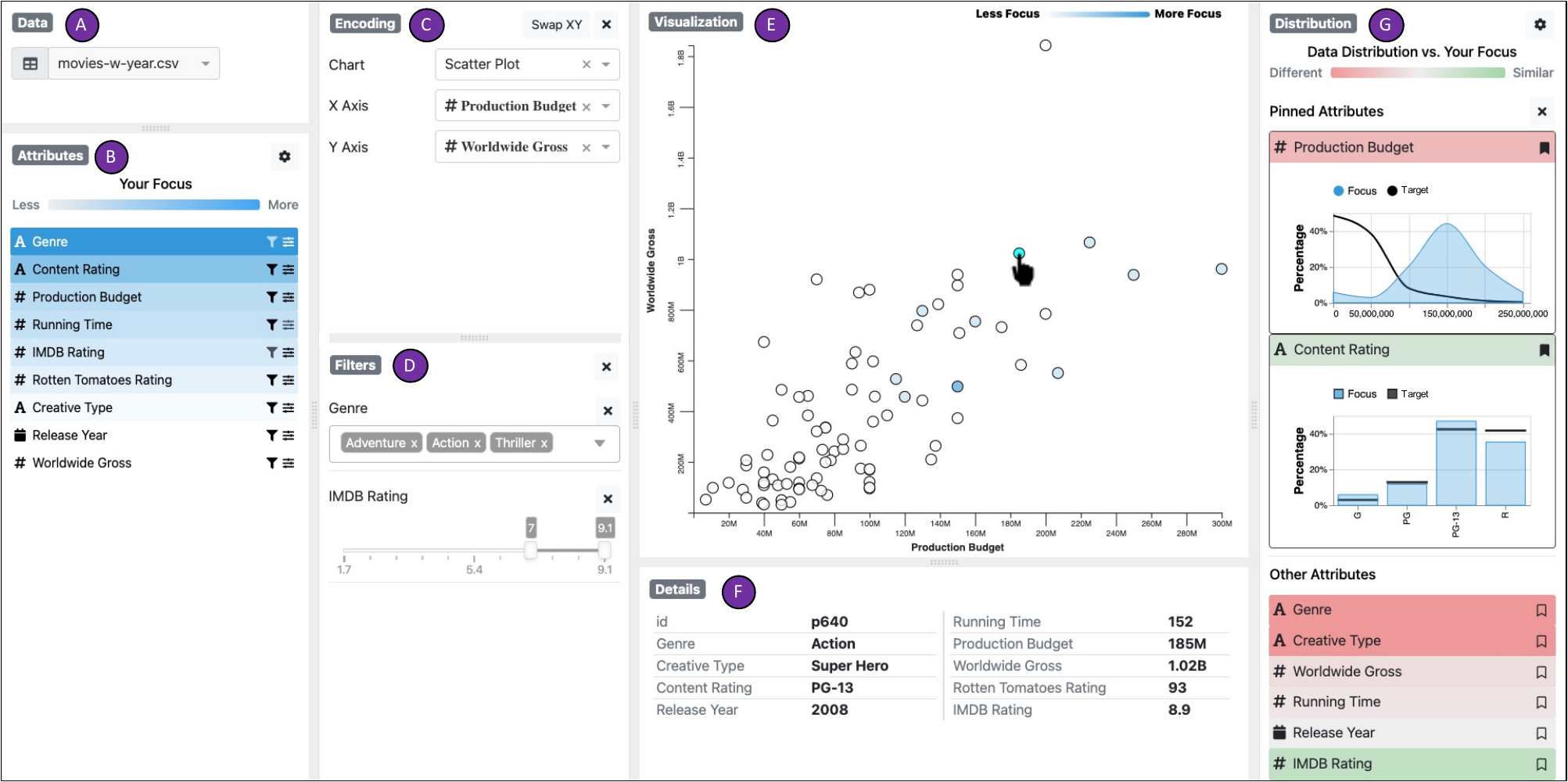}
  \caption{\revise{The \app user interface includes traditional visual data analysis functions (\textbf{(A)} \textbf{Data Panel}, \textbf{(B)} \textbf{Attributes Panel}, \textbf{(C)} \textbf{Encoding Panel}, \textbf{(D)} \textbf{Filter Panel}), and shows analytic behavior as in-situ and ex-situ interaction traces in the \textbf{(B)} \textbf{Attributes Panel}, \textbf{(E)} \textbf{Visualization Canvas}, \textbf{(F)} \textbf{Details View}, and \textbf{(G)} \textbf{Distribution Panel} as the relation between a target distribution (e.g., the underlying data) and the user's \revise{analytic behavior}}.
  }
  \label{fig:teaser}
}

\vgtcinsertpkg

\begin{document}


\firstsection{Introduction}

\maketitle

Visualizations take advantage of people's \revise{perception} 
to facilitate intuitive\revise{ly} understanding \remove{of }data.
Interactive features of visualizations become \remove{increasingly }critical when considering complex data, allowing people to progressively refine visual representations of data.
While it can\remove{facilitate agency and} aid in comprehension of large and complex data, certain patterns of interactivity \remove{with data }can signal insular data analysis practices. 
Users may be unknowingly stuck inside an ``echo chamber'', where their own unconscious biases may lead them to attend to certain parts of the data while neglecting others. 

Unconscious biases can take many forms, some of which are relatively innocuous (i.e., personal preference for a particular genre of movie) while others can lead to costly incorrect decisions or engender harmful societal stereotypes.
For instance, dark-skinned people are denied parole (racial bias)~\cite{angwinPropublica}, women experience a variety of barriers to equity in C-suite promotions (gender bias)~\cite{jordan2007gender}, ailing but younger people are denied optimal treatment (age bias)~\cite{sauerborn1996age}, etc. 
Even when these factors are not actively considered, they can implicitly affect the way people examine and process new information~\cite{greenwald1998measuring}, \revise{and hence their analytic behaviors}.

Apart from implicit biases and stereotypes, there are \remove{a number of }other cognitive and perceptual biases that also influence \revise{people's analytic behaviors.} 
Cognitive biases, in particular, describe systematic errors that can result from the use of ``fast and frugal'' heuristics~\cite{gigerenzer2004fast} to make decisions. 
Several biases have been shown to extend to tasks involving visualizations for decision making (e.g.,~\cite{wall2019formative,cho2017,dimara2017attraction,valdez2018priming}). 
Yet, common visual data analysis tools such as Tableau and Microsoft Excel that help users see and understand their data do not report \revise{analytic behaviors that may correspond to such biases.} 
Therefore we ask: how much can understanding data analysis and decision\revise{-}making \revise{behaviors} reduce the potentially negative influences of \revise{potential} cognitive, perceptual, or societal biases, if users were simply more aware of these often unconscious factors?

We present \app, an analysis tool that visualizes interaction history with data (i.e., \textbf{interaction traces}~\cite{lrg2021wall}) to increase \textbf{awareness} of potential \revise{interaction} biases that influence data analysis and decision making processes. 
Using \emph{in-situ} and \emph{ex-situ} visualization techniques, \app provides real-time feedback about a user's \textbf{\revise{analytic behavior}} for \revise{self-awareness and self-reflection to potentially change future course.} 
For example, \app remembers and highlights datapoints that have been previously examined in the same visualization (in-situ) and overlays the interacted datapoints on the underlying data distribution in a separate visualization (ex-situ) for comparison. 
Furthermore, \app allows users to configure a custom target distribution to reflect decision making goals (e.g., a university admissions committee in a computer science department may define an analysis target of 60\% female to promote increased gender diversity in the department, even if only 40\% of applicants are female). 
We posit that \app can improve behaviors exhibited during data exploration and decision-making to help mitigate the dangers of human \revise{interaction} biases affecting judgments and foster more transparent analysis processes.

The primary contributions of this work include \textbf{(1)} a technique to model users' \textbf{\revise{analytic behavior}} from interactions with the data, \textbf{(2)} a visual analysis tool, \app (\textbf{\url{https://lumos-vis.github.io/}}), that implements \remove{showing}\revise{visualizing} \revise{analytic behavior} using interaction traces, \textbf{(3)} a series of scenarios that describe potential usage of \app, and \textbf{(4)} results from a user study to understand how \app helps users be more aware of their \revise{analytic behavior} during visual data analysis.

\section{Related Work}
\label{section:related-work}
\subsection{Graphical History}
We are limited by our memory's capacity to remember and track our prior interactions with data in both amount and decay~\cite{miller1994magical, liu2014effects} which creates a barrier to data exploration.
Analyzing prior interactions with data in a visualization is one form of analytic provenance~\cite{ragan2015characterizing,north2011analytic} often used to infer about one's analysis process (beyond analytic outcomes).
Visualizing one's prior interactions can take many forms and lead to shifts in a user's analysis behavior. 
For instance, when users' prior interactions with charts or data points are encoded (e.g., analogous to coloring previously visited hyperlinks on a webpage purple), people tend to interact with more of the data~\cite{feng2016hindsight}.
Similarly, when exploration history is shown in interactive network visualizations, users report inspiration for conducting further analysis and greater recall of their prior explorations~\cite{dunne2012graphtrail}.
Cleverly designed histories can also help users maintain contextual awareness of previously visited data when distortions are applied that would otherwise make contextual awareness a challenge (e.g., fisheye lens)~\cite{skopik2005improving}. 
More generally, Heer et al. summarize a design space of mechanisms for displaying interaction histories~\cite{heer2008graphical}.
Given ongoing recognition within the visualization community that interaction histories can serve a wide range of purposes, recent tooling has been developed to support provenance tracking~\cite{cutler2020trrack}.

Graphical traces of user interactions have also been utilized in collaborative visualization settings, e.g., to facilitate coordination of multiple users by showing current selections and interactions as ``coverage'' of the data~\cite{sarvghad2015exploiting,badam2017supporting} and in personalized integrated development environments (IDEs), e.g., Footsteps for VSCode\footnote{https://marketplace.visualstudio.com/items?itemName=Wattenberger.footsteps} highlights the lines of code as they are edited by the user.
Similarly, showing social information ``scents'' on data visualization widgets (e.g., representing others' interactions with radio buttons, sliders, etc.) leads users to make substantially more unique discoveries in the data~\cite{willett2007scented}.

Outside of the visualization domain, traces of prior \remove{focus or }interactions have long been applied in HCI contexts, including revisiting common regions of a page using scrollbar history~\cite{alexander2009revisiting}, tracking interactions with documents by authors and readers~\cite{hill1992edit}, facilitating groupware coordination~\cite{gutwin2002traces}, tracking user focus while browsing a webpage using eye-tracking\cite{nielsen2010eyetracking} and mouse-tracking\cite{arroyo2006usability} gear, etc.
We take inspiration from this lineage of graphical representations of interaction history in our technique, \textbf{interaction traces}, expanded from~\cite{lrg2021wall} and described further in the next section.

\subsection{Modeling User \revise{Behavior}}
\label{sec:focus}

Several metrics have been proposed in data visualization to characterize behaviors during data analysis with visualizations. 
For instance, Feng, Peck, and Harrison propose metrics to quantify \emph{exploration uniqueness} and \emph{exploration pacing} as users interact with points in a scatterplot~\cite{feng2019patterns}.
Ottley, Garnett, and Wan use a hidden Markov model to capture user attention to predict clicks in a visualization~\cite{ottley2019follow}.
Perhaps most relevant to the present work, Gotz, Sun, and Cao model and visualize the provenance of how a user's subset selections of the data differ from the dataset as a whole~\cite{gotz2016adaptive}. 
We similarly model and visualize deviations of user behavior against a baseline; however, rather than focusing on explicit subset selections, we utilize a breadth of user interactions (including clicks, hovers, visualization configurations, filter configurations, etc.) as signals in our \revise{analytic behavior} model. 

Zhou et al. introduce a formal model of \emph{focus} based on user interactions defined by (1) type of action and (2) focus of the action in the form of an additive model~\cite{zhou2021modeling}. 
Wall et al. define metrics for quantifying \emph{bias}, similarly based on (1) type of interaction and (2) object of interaction; however, these metrics differ in that they compare the observed focus to a baseline of ``unbiased'' behavior~\cite{wall2017warning}.
While these metrics have been used to quantify bias (e.g., anchoring), they are also more generally used to capture deviations from uniform behavior~\cite{wall2019formative,wall2019markov}.
Because it compares user behavior to a flexibly defined baseline, in this work we utilize the metrics by Wall et al. to model \revise{user behavior}~\cite{wall2017warning}.


\section{Lumos}
\label{section:app}
In this section we first define the terminologies used in the paper, enumerate our design goals, and describe the \app user interface\remove{ along with brief implementation details}. 

\bpstart{Interaction Logs.}Telemetry data of a user's interactions (e.g., hovers) with the user interface elements (e.g., datapoints on a scatterplot).


\remove{\bpstart{Analytic Focus.}User's attention on data (e.g., distribution of interacted datapoints and attributes) modeled from the interaction logs.}

\bpstart{Analytic Behavior Model.}\revise{A model that quantifies user's behavior (set of actions) from interaction logs (e.g., computing the distribution of interacted datapoints and attributes).}


\bpstart{Interaction Traces.}Visual feedback of the user's \revise{analytic behavior} in the user interface; this is shown by adding visual scents in two ways: \emph{in-situ} (at the place of interaction) and \emph{ex-situ} (in an external view).

\bpstart{Awareness.}Knowing what is going on \cite{endsley1995toward} or what has been done and found during the exploration process to perform effective reasoning\cite{shrinivasan2008supporting}; in this context, knowledge gained from inspecting one's \revise{analytic behavior} via interaction traces.

\subsection{Design Goals}
Our development of \app was driven by \textbf{five} key design goals. We compiled these goals based on a combination of similar prior visual analysis tools \cite{feng2016hindsight, sarvghad2016visualizing, gotz2016adaptive}, formative feedback from pilot studies, and our own hypotheses with respect to usability.

\bpstart{DG1. Capture and present \revise{analytic behavior with} attributes.}
Overemphasis (or underemphasis) on specific attributes during data exploration may lead to unconscious biases (e.g., not interacting with a Gender attribute may practically result in a bias towards men if the dataset has more men than women). This goal translates to capturing user interactions with \emph{attributes}, \revise{modeling} \revise{analytic behavior}, and showing interaction traces to increase awareness to influence changes in subsequent interactions.

\bpstart{DG2. Capture and present \revise{analytic behavior with} datapoints.}
Overemphasis (or underemphasis) can also occur on specific values of data (e.g., interacting mostly with a few top candidates for university admissions may come at the expense of neglecting other candidates). \revise{This translates to the same goal as \textbf{DG1} but at the datapoint-level}.

\bpstart{DG3. Facilitate configuring different target distributions.} 
Determining overemphasis (or underemphasis) on specific attributes or data requires comparing a user's \revise{analytic behavior} with a known target distribution (e.g., the underlying data) as a baseline. However, different domains, tasks, attributes, or social norms may call for different target distributions. This goal translates to allowing users to configure different target distributions to suit their requirements.

\bpstart{DG4. Facilitate comparison between \revise{analytic behavior} and a baseline distribution.}
This goal translates to visualizing the user's \revise{analytic behavior} and the configured target distribution and quantifying the difference (or similarity) between the two distributions.

\bpstart{DG5. Facilitate visual data exploration while showing awareness.} 
This goal ensures that visual data exploration usability is not sacrificed by the added awareness visualization techniques. 


\subsection{\revise{Quantifying} Analytic \revise{Behavior}}
We quantify \emph{\revise{analytic behavior}} using (1) the attribute distribution (AD) metric~\cite{wall2017warning} and (2) relative frequency of interactions with data and attributes. 
The AD metric, along a scale from 0 (no bias) to 1 (high bias), characterizes how a user's interactive behavior deviates from expected behavior. 
By default, the system chooses a \textbf{proportional} baseline of expected behavior, wherein interactions with any given data point are equally likely, reflecting the true underlying distributions of attributes in the dataset. 
For example, if a user interacts primarily with PG-13 movies among a dataset of movies that contains predominantly G-rated movies, the AD metric for the Content Rating attribute will be high (more emphasis).
If the user instead spent more time interacting with G-rated movies, proportional to the distributions in the dataset, the AD metric value for Content Rating would be low (less emphasis). 
Alternative baselines can be set using \app (Section~\ref{sec:differentbaselines}).

\subsection{User Interface}
The \app user interface consists of the following views:


\begin{enumerate}[label=(\Alph*),nosep]
    \item \bpstart{Data Panel} shows the currently loaded dataset.
    \item \bpstart{Attribute Panel} shows a list of attributes in the dataset along with their datatypes: \{Nominal (N; \faFont), Quantitative (Q; \faHashtag), Temporal (T; \faCalendar)\} and buttons to apply a filter (\faFilter). 
    \item \bpstart{Encoding Panel} shows UI controls (dropdowns) to create visualizations by specifiying different encodings: \{Chart Type, X Axis, Y Axis, Aggregation\}.\remove{\textbf{(DG5)} summarized in Figure~\ref{tab:vis-matrix}.}
    \revise{\app currently supports \emph{four} visualization types: \emph{\{scatter plot, strip plot, bar chart, line chart\}} and \emph{five} aggregation types: \emph{\{count, sum, minimum, maximum, average\}} depending on the attribute data type combinations.} 
   \item \bpstart{Filter Panel} shows UI controls (range sliders for \{Q, T\} and multiselect-dropdowns for \{N\} attributes) to filter data. Filters can be added by clicking on \faFilter~ in the Attribute Panel (B) \textbf{(DG5)}.
    \item \bpstart{Visualization Canvas} renders the visualization based on the Encoding (C) and Filter (D) Panel specifications.
    \item \bpstart{Details View} shows additional information when the visualization elements (e.g., point, bar, strip) in (E) are interacted with. Hovering on a single datapoint (e.g., a strip in a strip plot) shows a list of all attribute values for the given datapoint (Figure~\ref{fig:teaser}F). Hovering on an aggregation of datapoints (e.g., a bar of a bar chart) shows a table of all datapoints that belong to that aggregation (the bar) with attributes as columns and values as rows (Figure~\ref{fig:details-view-agg}).
    \item \bpstart{Distribution Panel} shows a list of attribute cards similar to the Attribute Panel where clicking on a card toggles open/close a visualization that overlays user's \revise{interaction traces} on datapoints (blue area) on the target distribution (black curve) \textbf{(DG3)}. \remove{Clicking on \faBookmark~keeps the attribute open. }
\end{enumerate}


\begin{figure}
    \centering
    \setlength{\belowcaptionskip}{-12pt}
    \includegraphics[width=\columnwidth]{../figures/details-view-agg.pdf}
    \caption{\textbf{Details View (Aggregate).} Hovering on an aggregate visual element in the \emph{Visualization Canvas} (e.g., a bar in a bar chart) updates the Details View with a table of movies with \attr{Release Year}=2008; darker blue=\revise{greater emphasis toward that} datapoint.}
    \label{fig:details-view-agg}
\end{figure}

In addition to these features, \app supports additional customization options (\faCog): (1) \emph{Sort By} to sort attributes in the Attribute panel by their \{Order-in-Dataset (default), Name, Datatype, Focus\} (2) \emph{Color Mode} to determine the normalization strategy to compute the interaction frequency: \emph{Relative} (default) divides each value by the maximum of all values; \emph{Absolute} divides each value by the sum of all values; \emph{Binary} treats no interactions=\texttt{0} and at least one interaction=\texttt{1}, (3) \emph{Focus Mode} that determines the Y-encoding in the Distribution Panel visualizations: \emph{Percentage} (default) shows the percentage of Focus and Target distributions whereas \emph{Raw} shows the absolute counts; and (4) \emph{Color Scale} to choose different color scales (e.g., Sequential, Divergent).

Based on user interaction logs with attributes and datapoints, \app computes \revise{analytic behavior} and presents real-time interaction traces back to the user. \revise{Analytic behavior} and interaction traces are observed in multiple ways in the interface.

\subsubsection{In-situ Interaction Traces}

\begin{figure}
    \centering
    \setlength{\belowcaptionskip}{-12pt}
    \includegraphics[width=0.9\columnwidth]{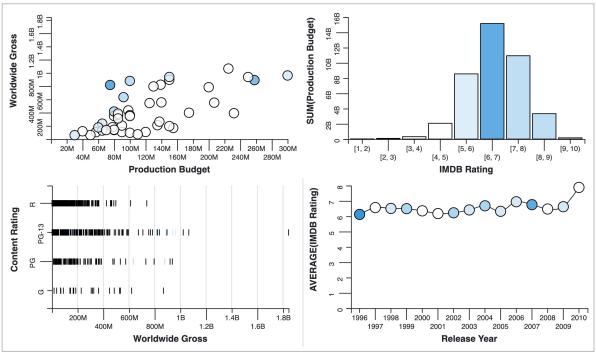}
    \caption{\app captures mouseover interactions on visualization elements (e.g., points, bars) and leaves traces in-situ by coloring (fill) them in shades of blue in proportion to the relative frequency of interactions.}
    \label{fig:insitu-awareness}
\end{figure}

\bpstart{Visualization Canvas.}
\app tracks user interactions with visual representations of datapoints (e.g., bars, lines, points, strips) and colors them on a white$\rightarrow$blue scale based on the relative frequency of interactions, e.g., dark blue color represents more interactions (Figure~\ref{fig:insitu-awareness}) and white represents no interactions \textbf{(DG 2)}. \app captures mouseover interactions as a proxy for \revise{modeling analytic behavior from interactions with} datapoints. \app employs a heuristic to ignore mouseovers that are active less than a 350 milliseconds threshold, regarded as random, accidental, or unintentional.

An interaction with a unit visualization (e.g., hovering on a point in a scatter plot of \attr{Running Time} and \attr{Worldwide Gross}) is handled differently than an interaction with an aggregate visualization (e.g., hovering on a bar showing average \attr{Running Time} of \emph{Action} movies). In the former scenario, \app treats it as one \emph{complete} interaction with the corresponding datapoint incrementing its interaction counter by \texttt{1}. In the latter scenario, \app treats the interaction as a set of \emph{partial} interactions with all constituent datapoints (e.g., all \emph{Action} movies), incrementing their corresponding interaction counters by \texttt{1/N} where N=number of constituent datapoints. 

\bpstart{Details View. }
When an aggregate visualization element (e.g, bar) is hovered on, the Details View below the chart shows a table with each datapoint. 
\app captures a mouseover on a table row, treats it as an interaction with the corresponding datapoint, and leaves an interaction trace by updating the table row's background color 
(Figure~\ref{fig:details-view-agg}) \textbf{(DG 2)}.

\bpstart{Attribute Panel. }
Like datapoints, \app also tracks user interactions with attributes. Each attribute card in the Attribute Panel is colored on a white$\rightarrow$blue scale based on the corresponding number of interactions (white=no interaction; darkest blue=most interactions).
\app captures attribute assignments to encodings (e.g., X, Y) and filters (e.g., Gender=Male) as a proxy for \revise{modeling analytic behavior from interactions with} attributes. These interactions are totaled and normalized relative to the most interacted attribute to determine the resultant shade of blue, e.g., in Figure~\ref{fig:teaser}B, \attr{Genre} has been interacted with most (dark blue) and \attr{Worldwide Gross} has not been interacted with at all \textbf{(DG 1)}. 

\subsubsection{Ex-situ Interaction Traces}

\app allows users to compare their \revise{analytic behavior} with a target distribution.
Attribute cards in the Distribution Panel are colored on a red$\rightarrow$grey$\rightarrow$green scale (Figure~\ref{fig:teaser}G) based on the difference between their respective \revise{analytic behavior} and underlying distributions (as quantified by Wall et al.'s AD metric~\cite{wall2017warning}, Section~\ref{sec:focus}) \textbf{(DG 4)}. 
In this evaluation, we set the target distribution to the underlying data. A red background indicates that the user's \revise{analytic behavior} is different from the target distribution (green background indicates similarity).
For example, inspecting the visualization for \attr{Production Budget} shows the \revise{analytic behavior} peaking around \texttt{150M} (blue area) when most movies have a budget under \texttt{50M} (black curve); the magnitude of the deviation is high resulting in a red background. 
Similarly, the computed \revise{analytic behavior} (blue bars) on \attr{Content Rating} is more closely matching the underlying data (black strips) resulting in a green background. 

\subsubsection{Configuring Different Target Distributions}
\label{sec:differentbaselines}

\begin{figure}[t]
    \centering
    \setlength{\belowcaptionskip}{-12pt}
    \includegraphics[width=\columnwidth]{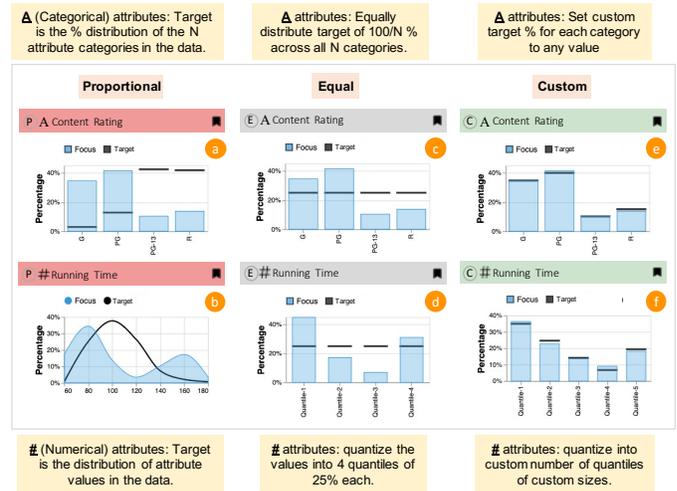}
    \caption{\app supports \revise{ex-situ interaction traces} for three modes of target distributions (\emph{Proportional}, \emph{Equal}, \emph{Custom}). \app presents these targets in the charts as black curves/strips along with \revise{user behavior} (blue area). \app also computes the difference between target and \remove{focus}\revise{observed behavior} and encodes it as the background color of the corresponding attribute card (red,grey,green colors where redder=more different; greener=more similar).}
    \label{fig:target-distribution-modes}
\end{figure}

Users can define target \revise{interactive analytic behavior} in multiple ways: (1) by \textbf{proportional} interactions across the various attribute distributions of the dataset, (2) by \textbf{equal} interactions across the categories of the dataset, and (3) by defining a \textbf{custom} target distribution of interactions across the data (\textbf{DG3}).
For example, consider a dataset of job applicants, where 50\% of applicants identify as male, 40\% of applicants identify as female, and 10\% of applicants identify as non-binary. 
A \textbf{proportional} baseline would define the target distribution of interactive behavior such that 50\% of interactions should be with male applicants, 40\% with female applicants, and 10\% with non-binary applicants, while an \textbf{equal} baseline would set the target distribution of interactions with 33.3\% male applicants, 33.3\% female applicants, and 33.3\% non-binary applicants. 
If, for instance, diversity is a target in filling this particular role, then a \textbf{custom} baseline might be set, where the target interaction distribution is 40\% female, 40\% non-binary, and 20\% male applicants.
Figure~\ref{fig:target-distribution-modes} summarizes these settings in the context of a movies dataset for the Content Rating attribute, and shows (in blue) how the user's actual \remove{interactions}\revise{analytic behaviors} compare to the target. 
\remove{Section \ref{subsection:customtargetdistribution} describes an example usage scenario demonstrating the application of \textbf{equal} and \textbf{custom} target distributions.}\revise{Users can configure \textbf{proportional}, \textbf{equal}, or \textbf{custom} target distributions per attribute in \app. In the \textbf{custom} mode for categorical attributes, users are presented with an interactive bar chart where they can drag individual bars (each representing a category) to their desired relative weights. For quantitative attributes, users can sketch a target distribution by clicking (to add new quantiles) and dragging points in the presented interactive histogram.}
While \app{} supports all three target distribution variations, our study (Section~\ref{section:evaluation}) utilizes a \textbf{proportional} configuration by default. 
In this evaluation, we focus on obtaining qualitative feedback from users about \emph{awareness of \revise{analytic behavior}} and defer the study of users defining target distributions to future work.

\subsubsection{Feedback from Pilot Studies}
The aforementioned \app interface and interaction design was realized after incorporating feedback from pilot studies with \texttt{five} participants. 
These helped us fix bugs, add enhancements (e.g., enable resizing the panels), reduce errors and slips (e.g., only showing the Aggregate encoding when applicable), and guide the design choices in the subsequent user study (e.g., we decided to remove the movie \attr{Title} from the dataset as participants reported that their experiential knowledge stimulated saliency biases \revise{causing}  
an exploration that neglected the underlying data almost entirely).

\remove{\subsection{Implementation}
The \app frontend is developed in Angular and interfaces with a Python-SocketIO server. User interactions are logged in the frontend and sent to the server which computes the \revise{analytic behavior model}
and relays them back to the frontend, all in real-time.}

\section{Example Usage Scenarios}
\label{section:scenarios}
\remove{In the following scenarios, we illustrate how \app can help users be more aware of their visual data analyses and potentially influence them to act on overemphasized or underemphasized aspects of data.}

\begin{figure*}[t]
    \centering
    \includegraphics[width=0.9\textwidth]{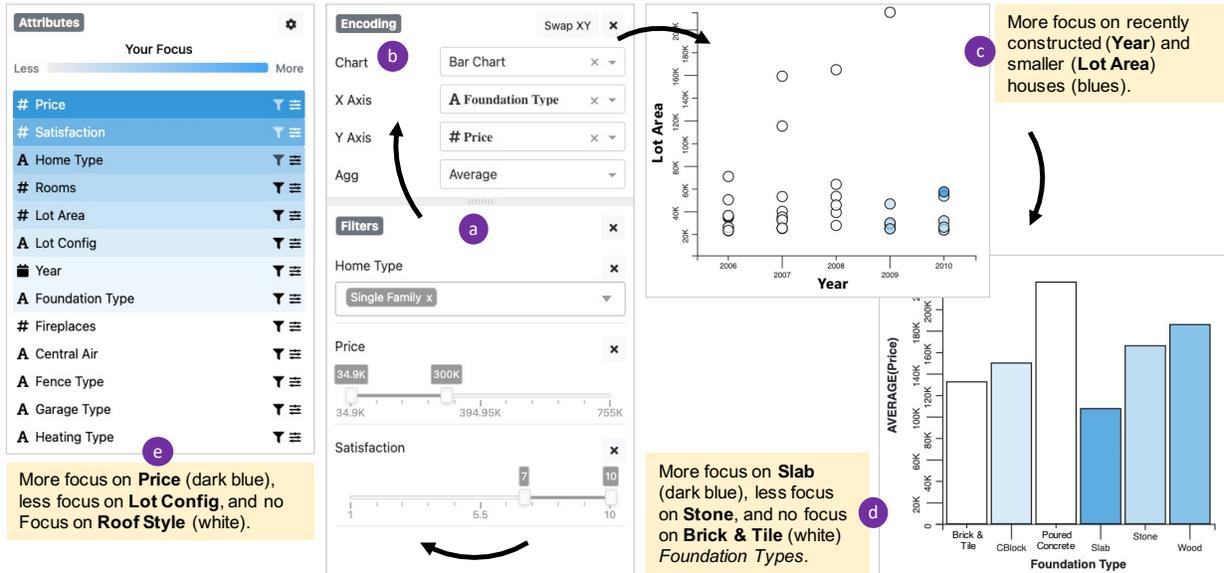}
    \caption{Scenario 1: \app helping \usera be more aware of their data analyses.}
    \label{fig:scenario-1}
\end{figure*}

\subsection{Scenario 1: Increasing awareness of \revise{analytic behavior}}
Assume \usera is looking for a new home and is exploring a housing dataset in \app (Figure~\ref{fig:scenario-1}). After acquainting themselves with the attributes, they apply three filters that match their criteria: \{\emph{Home Type=Single Family}; \emph{Price $\leq$ \$300K}; \emph{Satisfaction $\geq$ 7}\} (Figure~\ref{fig:scenario-1}a). Then, they create different visualizations by specifying encodings (Chart Type, X, Y, and Aggregation) in the Encoding panel (Figure~\ref{fig:scenario-1}b).

While interacting with these different visualizations, they observe visualization elements (e.g., bars, points) changing colors to different shades of blue. For example, in the scatterplot configuration with \attr{Lot Area} (Y axis) and \attr{Year} (X axis) (Figure~\ref{fig:scenario-1}c), \usera observes their focus has been on smaller (\emph{Lot Area $\le$ 60K}) and more recently constructed (\emph{2009 $\le$ Year $\le$ 2010}) homes. Similarly, in the barchart configuration with \attr{Foundation Type} (X axis) and \attr{Average(Price)} (Y axis) (Figure~\ref{fig:scenario-1}d), they observe they have not focused on \texttt{two} types of \attr{Foundation Types}: \{\emph{Brick \& Tile}, \emph{Poured Concrete}\} (white).

During their analyses, they also observe different shades of blue in the Attributes Panel (Figure~\ref{fig:scenario-1}e) inferring they have not focused on all attributes equally, e.g., they focused more on \attr{Price} and \attr{Satisfaction} (darker blues), not so much on \attr{Year} and \attr{Foundation Type} (light blues) and not-at-all on \attr{Fireplaces} and \attr{Heating Type} (white). 

After acknowledging that they did focus on the blue attributes, they start inspecting the \texttt{five} white attributes. They state they do not care about \{\attr{Lot Config} and \attr{Fence Type}\} but regret not focusing on the other \texttt{three} attributes (\attr{Heating Type}, \attr{Fireplaces}, \attr{Central Air}) associated with climate control as the city faces severe winters. Accordingly, they apply new filters and encodings and continue their analyses.

In this way, \app helped \usera in house-hunting by making them more aware of their \revise{analytic behavior with}\remove{focus on} data and attributes\remove{during their analyses}.

\begin{figure}[t]
    \centering
    \setlength{\belowcaptionskip}{-12pt}
    \includegraphics[width=\columnwidth]{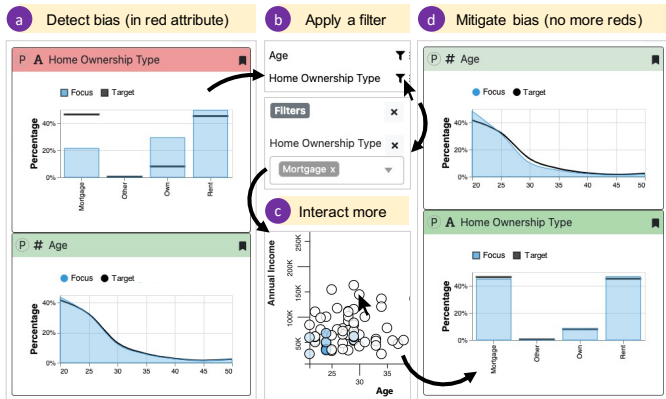}
    \caption{Scenario 2: \app helping \userb mitigate \remove{bias}\revise{biased analytic behavior} within the \attr{Home Ownership Type} attribute categories.}
    \label{fig:scenario-2}
\end{figure}

\subsection{Scenario 2: Mitigating \revise{biased analytic behavior}}
\userb, a loan officer, is using \app to analyze loan applications to determine credit-worthiness. After exploring the dataset for a while, they observe a red \attr{Home Ownership Type} attribute card and a green \attr{Age} attribute card (Figure~\ref{fig:scenario-2}a) in the Distribution Panel. They express happiness at not exhibiting any age bias but are concerned that their interactions with \attr{Home Ownership Type} significantly deviate from the underlying data (target) distribution. They click on the card to toggle it open and begin inspecting the visualization. They observe they have unknowingly overemphasized on \emph{Own} and \emph{Rent} and underemphasized on \emph{Mortgage} \attr{Home Ownership Type}. 

Willing to correct their \remove{unconscious bias
}\revise{behavior}, they apply a (reverse) filter: \{Home Ownership Type=\emph{Mortgage}\} (Figure~\ref{fig:scenario-2}b) and analyze a few previously unconsidered (white) points (Figure~\ref{fig:scenario-2}c). They finally see a greener \attr{Home Ownership Type} card (Figure~\ref{fig:scenario-2}d) and are more content.

\section{Evaluation}
\label{section:evaluation}
\subsection{Study Design}

We conducted a between-subjects qualitative study with the aim of understanding how \app helps users increase awareness of their \revise{analytic behaviors}. The study protocol was reviewed and approved by our institutional review board.\remove{ In the following sections we describe the participants, detail the high-level procedure, and present and discuss our findings.}

We recruited 24 participants (Gender: 14 male, 8 female, 1 other, 1 preferred not to say; Age in years: $\mu$=24.04, $\sigma$=3.44, median=24, 1 preferred not to say). Participants included students, researchers, and industry professionals with diverse educational degrees (8 bachelors, 7 masters, 9 doctoral) \revise{all from a computing or related field (e.g., computer science, human-computer interaction, human-centered computing)} with a self-reported visualization literacy $\ge$ 3 (on a 5-point Likert scale). Sessions lasted approximately 1 hour and participants were each compensated with a \$15 Amazon gift card.

Due to the COVID-19 pandemic, we conducted the study remotely using online collaboration tools. 
The actual study was conducted over the Microsoft Teams (https://www.microsoft.com/en-us/microsoft-teams/group-chat-software) teleconferencing software. The experimenter provided participants access to the study environment by sharing their computer’s screen and granting input control.

Participants were randomly divided into either a \textbf{Control} or \textbf{Awareness} condition, which determined the version of the system they used for the study. \revise{Both conditions had roughly equal numbers of participants with bachelors, masters, and doctoral degrees (Awareness: bachelors=4, masters=4, doctoral=4; Control: bachelors=4, masters=3, doctoral=5).} 
Participants in the Control condition \emph{did not} see the Distribution Panel (along with the \emph{ex-situ} interaction traces) nor did they see the \emph{in-situ} interaction traces in the Visualization Canvas, Details View, and Attribute Panel. 
We pre-configured the target distribution to \emph{Proportional} and hid the Settings Panel for both conditions.

Each participant first electronically signed a consent form. Then, depending on the study condition, they saw a 2 minute (Control) or 4 minute (Awareness) video tutorial that demonstrated the features of \app. Participants were then asked to perform a practice task using a dataset of cars to interact with and get acquainted with \app. After practice, we tasked participants to \emph{``analyze a dataset of movies to recommend the characteristics of movies that a movie production company should make next''}. The dataset consisted of \texttt{709} movies (rows) and \texttt{9} attributes (columns):  \attr{Production Budget}~(\faHashtag), \attr{Worldwide Gross}~(\faHashtag), \attr{Running Time}~(\faHashtag), \attr{IMDB Rating}~(\faHashtag), \attr{Rotten Tomatoes Rating}~(\faHashtag), \attr{Release Year}~(\faCalendar), \attr{Content Rating}~(\faFont), \attr{Genre}~(\faFont), and \attr{Creative Type}~(\faFont). Participants were encouraged to think aloud during the task. Throughout the task, their interactions were recorded.
The study ended with a debriefing in which participants discussed their overall experience with the system, provided suggestions for improvements, and completed a questionnaire to rate the usefulness of \app features. 
All sessions were screen- and audio-recorded for qualitative analysis.

\begin{figure*}[t]
    \centering
    \includegraphics[width=\linewidth]{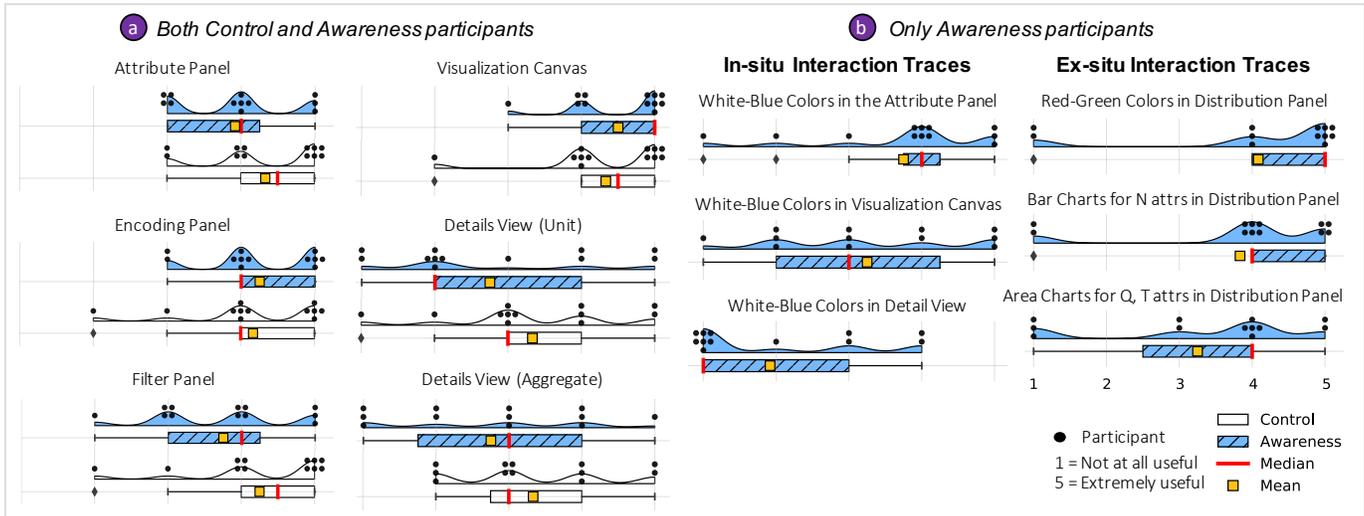}
    \caption{Summary of usefulness scores of all \app features as reported by participants in the post-study questionnaire\revise{, as RainCloudPlots~\cite{allen2019raincloud}}.}
    \label{fig:usability-scores}
\end{figure*}

\begin{figure*}[t]
    \centering
    \setlength{\belowcaptionskip}{-12pt}
    \includegraphics[width=\linewidth]{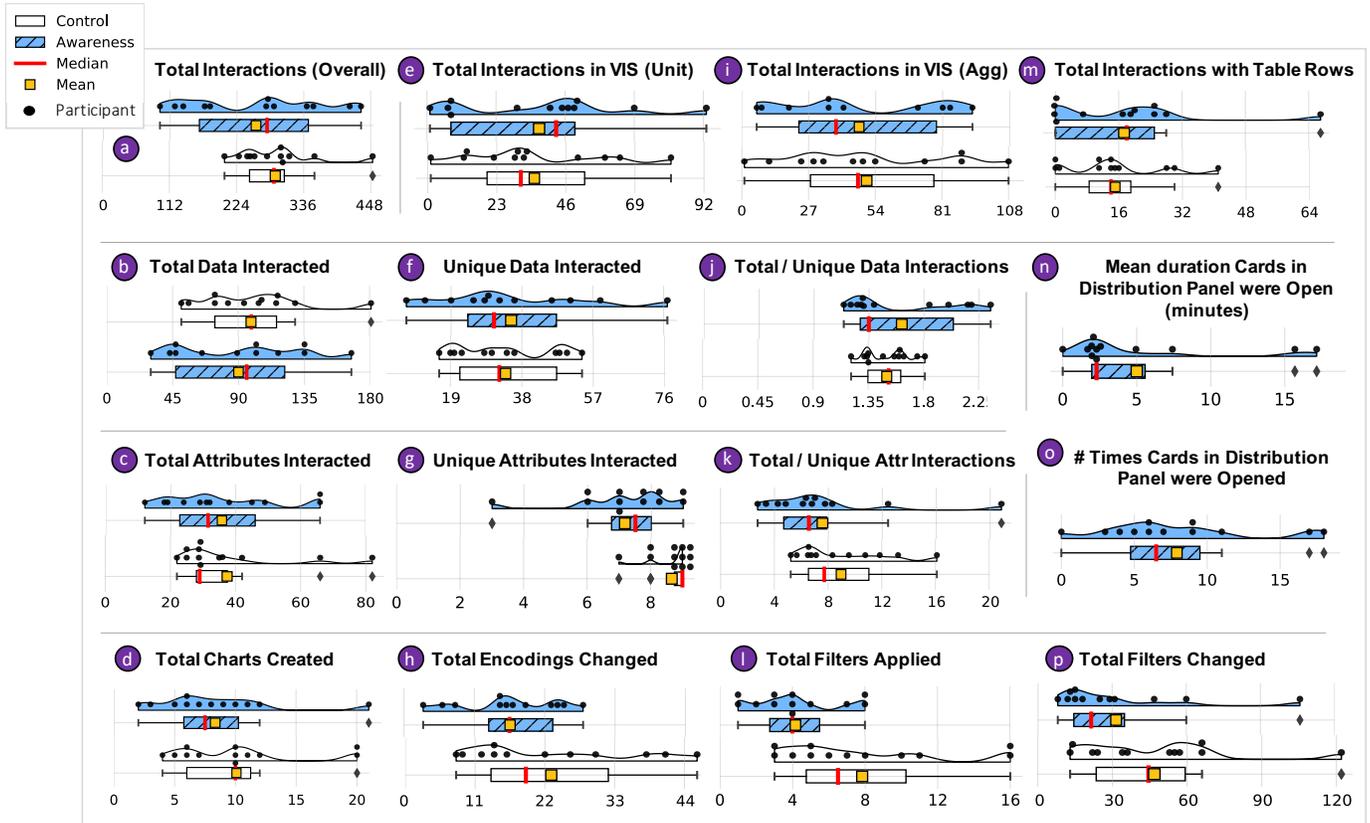}
    \caption{Summary of quantitative findings across the two experimental conditions\revise{, as RainCloudPlots~\cite{allen2019raincloud}. VIS=Visualization; Attr=Attribute.}}
    \label{fig:quantitative-results}
\end{figure*}

\subsection{Hypotheses}

We structure our study analysis according to the following hypotheses:
\begin{enumerate}[nosep]
\item [\textbf{H1}] Interaction traces will increase awareness of \revise{analytic behavior}.
\item [\textbf{H2}] There will be differences in interactive behaviors of Awareness v. Control participants (as measured by bias metrics~\cite{wall2017warning}, differences in use of filters, and number of charts created).
\item [\textbf{H3}] Participants will find the ex-situ awareness features to have greater utility than in-situ awareness features. 
\item [\textbf{H4}] Participants in the Awareness condition will react to interaction traces in ways to reduce potential \remove{biases}\revise{biased analytic behaviors}. 
\end{enumerate}

\subsection{Results}
Below, we present findings from the study and discuss them in context of participant feedback. \pa{01}-\pa{12} and \pc{13}-\pc{24} refer to the \texttt{24} participants in Awareness and Control conditions, respectively. Participant quotes and moments of awareness were coded and categorized using affinity diagramming\remove{ by the study administrators}. \revise{One administrator came up with an initial set of categories that were then refined during iterations with three other administrators until a consensus was reached. 
}

\subsubsection{General Feedback} Overall participant feedback was positive with \pa{01} commenting, \emph{``[they] haven’t seen many things like [\app{}] before...really good technique''}. \pa{09} mentioned that \emph{``[\app{}] can remove the internal bias of things users think are of the most interest''}. Participants found \emph{``the ability to keep track of [their] provenance, interaction history [as] interesting''} (\pa{08}) and \emph{``communicating it back [how they are doing] as something [they] would use in [their] tools''} (\pa{07}). \pa{09} found the Distribution panel \emph{``a great idea to show users what their focus was''} and \pa{05} found it \emph{``very helpful as [they] don't need to create visualizations in the Vis panel for each attribute to see [their] distributions''}. \pa{05} suggested \emph{``integrating this tool into existing tools such as Tableau [as] they don’t have a feature that tells [them] what attributes haven’t been used yet''} (\pa{12}). \pa{07} suggested \emph{``there are lots of use-cases for this technique in journalism and social media, e.g., you have only looked at Trump's negative tweets, but what about Biden's?''}. \revise{Two} participants found the Distribution Panel less useful as they \emph{``didn't know exactly what to do about the [red-green] cards''} (\pa{5}) or felt it \emph{``out of focus''} on the right side of the screen (\pa{10}).

\subsubsection{Usefulness Scores: \app user interface}
Figure~\ref{fig:usability-scores}a summarizes \app's usefulness scores (\emph{1=Not useful at all; 5=Extremely useful}) as reported in the post-study questionnaire:

\bpstart{Attribute Panel.} \revise{Participants generally found the attribute list useful ($\geq$3 out of 5, median$_A$=4, median$_C$=4.5) along with \emph{``their data types''} (\pc{21}) \emph{``unlike, e.g., Excel where they aren’t always on-screen''} (\pc{17}).} 

\bpstart{Encoding Panel.} \revise{\texttt{23}} participants found the Encoding Panel to be useful (median$_A$=4, median$_C$=4, \emph{``it is standard in a good way''}-\pc{21}) except \pc{23} who found it \emph{``only slightly useful''}. \revise{Four} participants noted that the system messages to fix incorrect encodings were intuitive and helpful (\emph{``it is sometimes hard to know what's wrong in Tableau''}-\pa{05}) but \revise{two} found them confusing and suggested the app \emph{``prevent [them] from choosing incorrect encodings''} (\pc{18}) by \emph{``filtering out the chart types that are not allowed''} (\pa{11}). \revise{Five} participants also suggested additional features (\emph{``add color as a third encoding''} - \pc{24}) and enhancements (\emph{``support drag-drop attributes to the Encoding Panel''} - \pc{\{17,18,19\}}, \emph{``support text entry for the dropdowns''} - \pa{12}).

\bpstart{Filter Panel.} Participants utilized filters (median$_A$=4, median$_C$=4.5) \emph{``to remove outliers and to confirm hypotheses about the data''} (\pc{22}), to see the \revise{different values for a categorical} attribute (\pc{19}, \pc{23}), and to mitigate any unconscious \remove{bias}\revise{biased analytic behavior} (e.g., \emph{``Comedy and Drama are high percentage in the dataset, and I haven’t interacted with them at all, so it might be worth my time to look at them.''}-\pa{09}). \revise{\pa{10}} did not utilize filters as they were \emph{``being more exploratory with [their] analysis and if [they] wanted to look at finer details, [they] would have used more filters.''} \revise{Three} participants also requested enhancements to \emph{``specify precise inputs [for Q,T values]''} (\pc{22}), \emph{``allow hover on a value in the categorical filter and highlight in the Visualization''} (\pa{09}), and \emph{``support select- and deselect- all for categorical values''} (\pa{11}).

\bpstart{Visualization Canvas.}
\revise{23} participants found the Visualization Canvas useful (median$_A$=5, median$_C$=4.5), utilizing it to \emph{``observe patterns and outliers''} (\pa{10}), and \emph{``see the different categories and values in the attributes''} (\pc{14}). \revise{\pc{18}} \emph{``did not find it useful because a third attribute encoding, e.g., color was not supported''}.

\bpstart{Details View (Unit).} 
This view received mixed usefulness scores from our participants (median$_A$=2, median$_C$=3). \pc{17} \emph{``liked the Details portion and being able to hover over points for more details.''} \pc{19} and \pa{11} \emph{``didn’t find the Details view for single data points super useful''} because they wanted the name of the film to bring prior knowledge to the analysis and spark different hypotheses.\footnote{Note: the movie \attr{Title} was deliberately not shown in the Detail View to prevent personal experiences with data to influence the analysis.}

\bpstart{Details View (Agg).} 
This view also received mixed usefulness scores from our participants (median$_A$=3, median$_C$=3). \pc{18} found it to be \emph{``really useful because it is not apparent from the bar shape and size that some bars only have one point in them versus some bars having six or seven points''}. \pa{07} said \emph{``[they] don't really hover on things in e.g., a scatterplot but information shown on hovering a bar chart [the details view agg] was awesome because you showed individual data''}. Also, \emph{``it shows all information in one space.''} (\pc{20}, \pc{24}) and \emph{``could be useful for multivariate hypotheses''} (\pa{10}). \revise{Two} participants found it \emph{``hard to draw conclusions from lists of words and data''} (\pc{17}, \pc{19}) and preferred to see the information visually utilizing the Details View \emph{``only as a reference''} (\pc{19}). \revise{\pc{18}} utilized the Details View \emph{``because the visualization wasn’t as helpful''}.

\remove{\bpstart{Summary.} 
Comparing the distributions of scores (boxplots) for the aforementioned features (Figure 9a), Control participants scored the \{Attribute Panel, Filter Panel, and Details View\} higher than the Awareness participants whereas scores for the \{Encoding Panel, and Visualization Canvas\} are similar.}
\remove{We posit this was because Control participants had a lesser cognitive load than Awareness participants who, in addition to the Distribution Panel, also saw the other in-situ and ex-situ interaction traces. Also, Control participants did not have access to the ready-made attribute distribution charts leading them to rely more heavily on filters and visualizations.}


\subsubsection{Usefulness Scores: \app technique}
Participants in the Awareness condition also saw \emph{in-situ} and \emph{ex-situ} (Distribution Panel) interaction traces in the user interface. Figure~\ref{fig:usability-scores}b summarizes the usefulness scores (\emph{1=Not useful at all; 5=Extremely useful}) as reported in the post-study questionnaire:

\bpstart{Difference between \revise{analytic behavior} and target distribution.} \revise{Ten} participants found the difference between their \revise{analytic behavior} and the underlying data (red-green coloring in Distribution Panel) very useful (median$_A$=5) \emph{``giving a sense if I'm looking at the dataset in an unbiased way'' (\pa{06})}.

\bpstart{Ex-situ interaction traces in Distribution Panel.} The overall feedback on the \emph{ex-situ} interaction traces in the Distribution Panel was positive; participants found the bar charts for N attributes (median$_A$=4) more useful than the area charts for Q,T attributes (median$_A$=4) with \pa{06} nicely summarizing, \emph{``I can track and channel my focus based on discrete bar charts by applying a filter...but it is difficult to discretize and track a continuous [Q,T] attribute''}. 
For \revise{eight} users, these real-time traces helped increase awareness (\emph{``Geez, I haven't looked at Drama movies at all''}-\pa{07}), influencing them to interact differently (e.g., by creating a bar chart with \attr{Genre} to inspect movies of other potentially underemphasized genres) while \revise{two participants} either ignored them (\emph{``I never looked at the individual distributions of attributes''}-\pa{12}) or preferred to look at them after analysis \emph{``as the bars will be moving, and that’s distracting''} (\pa{09}).

\bpstart{In-situ interaction traces in Visualization Canvas.} \revise{There was mixed response to the \emph{in-situ} interaction traces (white-blue colors) in the Visualization Canvas (median$_A$=3)}. For \revise{\pa{06}}, they \emph{``helped in tracking visited points''} nudging them to interact with points that were not interacted yet, while for \revise{\pa{05}} they were confusing and distracting, nudging them to re-interact with them. \pa{08} did not want the colors to stay persistent but \emph{``be able to clear existing interactions and start a new session with a new set of model movies for comparison''}.

\bpstart{In-situ interaction traces in Details View.} 
\revise{Only four participants found the \emph{in-situ} interaction traces (white-blue colors) in the Details View to be useful (median$_A$=1)}.

\bpstart{In-situ interaction traces in Attribute Panel.} Participants generally found the \emph{in-situ} interaction traces (white-blue colors) in the Attribute Panel to be useful (median$_A$=4).
For \revise{\pa{05}}, they helped increase awareness of already-interacted attributes (\emph{``I see that I have spent a lot of time on Release Year so I’ll now see something else''}) while for \revise{\pa{12}}, they required some time to get acquainted with (\emph{``the coloring in the Attributes panel...I didn’t use it initially, and later on it hit me that I had this feature. Once I noticed it, it was very useful''}).

\bpstart{Summary.} Comparing the distributions of scores for the aforementioned features (Figure~\ref{fig:usability-scores}b), participants found the \emph{ex-situ} interaction traces more useful than the \emph{in-situ} interaction traces, supporting \textbf{H3}, consistent with experimental results from~\cite{lrg2021wall}. \revise{We believe this is because \emph{in-situ} interaction traces are always visible to a user whereas \emph{ex-situ} interaction traces can be used more on-demand without side-tracking the analysis task at hand. Furthermore, \emph{in-situ} traces block an otherwise common attribute encoding channel, \emph{color}, that can be undesirable for and cause inconvenience to some users.}

\subsubsection{Awareness Moments}
\bpstart{In-situ traces in the Attributes Panel.}
There were instances when Control participants expressed a need for tracking the already-interacted attributes. For example, we observed \pc{13} use hand gestures to recollect and count the attributes that they had already visited and \pc{14} exclaimed, \emph{``I hope I have interacted with all (attributes)''}. Awareness participants, on the other hand, saw the interaction traces and had several instances of \emph{awareness} during their respective analyses. \revise{Two} participants \emph{acknowledged} their choices (\emph{``I don’t think Release Year should matter too much, hence I am not interacting with it.''}-\pa{04}, \emph{``I don’t think Runnning Time is important to me''}-\pa{07}) while \revise{two} participants also suggested correcting future course via \emph{interaction} (\emph{``I see that I have spent a lot of time on Release Year so I’ll now see something else''}-\pa{05}, \emph{``[on seeing a white attribute bar] now I’m going to interact with Running Time''}-\pa{22}), also supporting \textbf{H4}. \revise{Two} participants reflected upon their choices after the study while they were answering questions pertaining to self-reported focus on individual attributes in the post-study questionnaire (\emph{``actually I forgot, had I remembered, it might have been interesting to not click on the same thing over and over.''}-\pa{03}, \emph{``I didn’t use the blue attributes panel but now that I see these questions, I would’ve seen them more''}-\pa{08}). These and the desire for awareness moments by Control participants validate \textbf{H1}.

\bpstart{In-situ traces in the Visualization Canvas and Details View.} \revise{Eight} participants found the \emph{in-situ} interaction traces in the visualization canvas to be useful; \revise{two participants} took some time to get acquainted with them (\emph{``very useful but I learnt about them slightly afterwards''}-\pa{01}, \emph{``I was initially confused but then over use I got used to them and found them useful in tracking visited points''}-\pa{06}). \pa{03} found the colors to be useful but questioned the technique because \emph{``if it is based on [me] hovering on a point again and again, it might not be 100\% correct.''} \pa{05} was distracted and \emph{``getting drawn to the visited points (instead of the white un-visited points).''} \revise{There was minimal commenting} on the \emph{in-situ} traces in the Details View but it led to some awareness for \pa{04} who hovered on an uninteracted (white) bar in a bar chart and observed \emph{``there aren't many blue rows which means I haven’t been focusing on it.''}

\bpstart{Ex-situ traces in the Distribution Panel.} 
There were multiple instances of awareness among participants (supporting \textbf{H4}).
\pa{04} verified the interactions traces by comparing it with ground truth (\emph{``distribution of my focus on Running Time (blue) is representative of the applied filter''}). 
\pa{05} reflected upon seeing three red attributes in the Distribution Panel and hypothesized that they were \emph{``just thinking aloud and exploring and will (now) follow a more targeted approach''}.
\pa{05} reflected upon seeing a red \attr{Content Rating} attribute (\emph{``Seems I didn’t interact with R-rated movies enough so this view nudges me towards those''}) and applied a filter to show only R-rated movies.
\pa{07} reflected upon the white \emph{Drama} category in the Distribution Panel and justified that they \emph{``didn’t care about Dramatization movies [...] who cares?''}  At one point, \pa{07} observed many red cards and exclaimed, \emph{``this is so biased but whatever.''} 
\pa{11} tried to correlate the effects of their interactions with different attributes (\emph{``I noticed that Adventure is representative of most values in Rotten Tomatoes Rating and IMDB Rating [...] This is because I mostly interacted with just Adventure movies and that caused those attributes [in the Distribution Panel] to be colored green''}).
\pa{10} did not use the Distribution Panel as they \emph{``didn't know how to use it in the context of what [they were] doing''}.

\subsubsection{Interactive Behaviors across Experimental Conditions}
Overall, Awareness participants exhibited more diverse behavior in terms of interactions with datapoints and attributes than Control (Figure~\ref{fig:quantitative-results}a; $\mu_A$=256, $\sigma_A$=115; $\mu_C$=289, $\sigma_C$=66), validating \textbf{H2}. 
\remove{We believe this is due to the increased cognitive processing load with the addition of the Distribution Panel and the interaction traces.}

\bpstart{Interactions with Datapoints.}
Figure~\ref{fig:quantitative-results}b shows the distribution of all interactions with datapoints across conditions ($\mu_A$=89.92, $\sigma_A$=43.83, $\mu_C$=98.75; $\sigma_C$=35.67). 
\revise{\pa{08}, on seeing the interaction traces, became fearful of skewing their interactions}. Both conditions interacted with a similar number of unique datapoints (Figure~\ref{fig:quantitative-results}f; $\mu_A$=35.08, $\sigma_A$=20.18, $\mu_C$=33.67; $\sigma_C$=13.36) indicating that the \emph{in-situ} blue colorings did not nudge participants to interact with more white points (in fact, \pa{05} felt drawn towards the blue points). Interestingly, this result seems different from the one found by Hindsight~\cite{feng2016hindsight}, where participants tended to interact with more unique points when data were colored based on prior interactions.
Figures~\ref{fig:quantitative-results}(e,i,m) show the distribution of interactions with (e) datapoints in unit visualizations ($\mu_A$=37.33, $\sigma_A$=27.93, $\mu_C$=35.50; $\sigma_C$=23.68), (i) aggregate visualizations ($\mu_A$=47.3, $\sigma_A$=32.53, $\mu_C$=50.5; $\sigma_C$=33.51), and (m) the Details View table rows ($\mu_A$=17.33, $\sigma_A$=19.11, $\mu_C$=15.08; $\sigma_C$=12.62).

\bpstart{Interactions with Attributes.}
Figure~\ref{fig:quantitative-results}(c,d) show the distribution of total ($\mu_A$=35.75, $\sigma_A$=17.8, $\mu_C$=37.42; $\sigma_C$=18.25) and unique interactions ($\mu_A$=7.17, $\sigma_A$=1.64, $\mu_C$=8.67, $\sigma_C$=0.65) with attributes across conditions, respectively. \revise{Four} awareness participants did not interact with certain attributes as they were not important to their analysis (\emph{``I don't think this [Running Time] is important to me''}-\pa{11}) while \revise{two others} tried to interact with more attributes to try and mitigate their \revise{biased analyic behaviors} with attributes (\pa{05}) and datapoints (\pa{07}).

\bpstart{Charts, Filters, and Encodings.}
On average, Control participants created more charts (Figure~\ref{fig:quantitative-results}d; $\mu_A$=8.33, $\sigma_A$=5.02, $\mu_C$=10.08, $\sigma_C$=5.28), changed more encodings (Figure~\ref{fig:quantitative-results}h; $\mu_A$=16.58, $\sigma_A$=7.86, $\mu_C$=23.00, $\sigma_C$=13.11), and applied (Figure~\ref{fig:quantitative-results}l; $\mu_A$=4.17, $\sigma_A$=2.44, $\mu_C$=7.83, $\sigma_C$=4.57)) and changed (Figure~\ref{fig:quantitative-results}p; $\mu_A$=31.58, $\sigma_A$=28.05, $\mu_C$=46.83, $\sigma_C$=30.50)) more filters than the Awareness participants. 


\bpstart{Cards in the Distribution Panel.}
\revise{Eleven} awareness participants utilized the Distribution Panel cards to check the data distribution (black curve) and/or compare their \revise{analytic behavior} with the underlying data. These participants opened these cards \revise{multiple times} at different points of their analysis (Figure~\ref{fig:quantitative-results}o, $\mu_A$=7.91 times, $\sigma_A$=5.35 times) and kept them open for different durations (Figure~\ref{fig:quantitative-results}n, $\mu_A$=5.00 minutes, $\sigma_A$=5.65 minutes). \pa{09} opened the cards but commented that they would prefer analyzing their performance \emph{after} the task was over. \revise{One} awareness participant did not utilize these cards as they were content with the red-green backgrounds of the closed cards (\pa{12}). \remove{Only \pa{06} utilized the pinning (\faBookmark) feature probably because the movies dataset had just nine attributes that fit nicely in the screen (\emph{``it will be useful when there are many attributes''}-\pa{06}).}

\subsubsection{Temporal analysis}
\begin{figure}[t]
    \centering
    \setlength{\belowcaptionskip}{-14pt}
    \includegraphics[width=0.9\columnwidth]{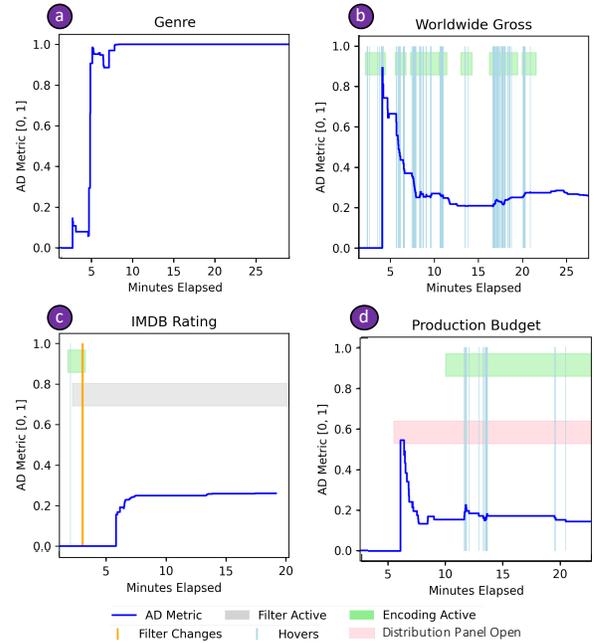}
    \caption{AD metric values over time for (a) \pc{16}, (b) \pc{13}, (c) \pa{09}, and (d) \pa{01} for specific attributes.}
    \label{fig:temporal}
\end{figure}

To understand the cadence of interactions, we plotted hover, filter, encoding, and Distribution Panel interactions over time, and superimposed the AD metric~\cite{wall2017warning} (representing analytic behavior) for each attribute. 
We found that for some participants who did not interact with these UI components, their \revise{analytic behavior} was skewed for specific attributes (as evidenced by high AD metric values for \attr{Genre} for \pc{16} in Figure~\ref{fig:temporal}a).
For others, interactions led to significant shifts in \revise{analytic behavior}. 
For instance, while \pc{13} used \attr{Worldwide Gross} as an encoding in a chart, their \revise{analytic behavior} toward the attribute became more proportional to the underlying data (Figure~\ref{fig:temporal}b).
While a filter was applied to IMDB Rating ($\ge$2.3 and $\le$9.1), \pa{09}'s \revise{analytic behavior} began to deviate from the underlying data as shown by increasing AD metric value (Figure~\ref{fig:temporal}c).
Lastly, while we found instances where inspecting the Distribution Panel and accordingly interacting with the visualization appear to reduce \revise{biased analytic behavior} (e.g., Figure~\ref{fig:temporal}d, \pa{01}), such instances were not abundant among our participants. 
\revise{Thus, temporal analysis provides minimal evidence to support \textbf{H4}, but interaction traces provided participants opportunities to reflect on their analytic behavior, judging how their process differs from the baseline, and if that difference is appropriate for their task.}


\section{Discussion}
\label{section:discussion}

\subsection{Using Color to visualize interaction traces}
\bpstart{Fun, Focus, Distraction, or Inconvenience?}
\revise{
For \pa{08}, \emph{``interacting [with scatterplot points] and seeing colors change was fun''} but for \pa{05}, it was a distraction as they were \emph{``getting drawn to the visited points [instead of the unvisited points]''}. \pa{09} noted \emph{``highlighting the same data point over and over again skews the distribution visualization, which could be an issue''} while \pa{08} became \emph{``careful to not interact with dots and skew their interaction trace''}. \pa{10} asserted \emph{``a hover does not imply I am interested in that point''} and altogether ignored the blue colorings while \pa{07} suggested capturing \emph{``mouse interactions based on proximity and not an exact hover''} \remove{as a proxy for focus}\revise{to compute analytic behavior}. Next, using color to encode interaction traces took away an encoding channel that could otherwise be used by an attribute. 
In fact, \pa{02} mentioned they \emph{``confused the blue colors with an [attribute] encoding''}. \pa{12} also asked if there was \emph{``a reason for having the coloring throughout the visualization?''} Furthermore, \pa{12} raised an accessibility concern noting, \emph{``making the tool [\app] color-blind safe would be really important''}. Thus, per our participants, using color to visualize interaction traces can be fun, cause a shift in emphasis, cause inconvenience, or be a distraction motivating the need to study alternate encoding approaches.}

\bpstart{If not Color, then what?}
\revise{While in this evaluation of \app, we studied the usage of color (shades of blue, red, and green) to encode interaction traces, there are other visual variables that can be modified to encode and convey the same information, e.g., \emph{stroke color}, \emph{stroke width}, \emph{size}, \emph{shape}, \emph{orientation}, \emph{opacity}, etc. It must be noted that some visual variables might work better for certain visualizations than others (e.g., modifying color works better for a scatterplot than a strip plot) and that some variables may not even be applicable for certain visualizations (e.g., it is difficult to modify the shape of a line chart). Exploring this design space will help derive guidelines for effective in-situ and ex-situ visualizations.}



\subsection{The role of target distributions}
\revise{
Target distributions in \app serve as a benchmark against which analytic behavior can be compared. For some tasks, meaningful target distributions may exist (e.g., forming committees with specific representation from certain groups). However, it may be harder to articulate a target distribution for other decision-making tasks, in which case standard baselines for comparison are more meaningful. \app allows users to modify these or use the data distribution as a default.}

\subsection{False positives in modeling analytic behavior}
\revise{The \app technique of presenting interaction traces can be subject to false positives. For example, users might intentionally not interact with an attribute because it is either not important or they are not interested in it. Labeling these as underemphasized may not be correct, as it was a conscious decision by users to ignore them. Furthermore, a categorical attribute, when encoded along one of the scatterplot axes, can lead to the formation of visual clusters that offload a cognitive task to a perceptual one, rendering that attribute's filter somewhat redundant. On the other hand, users can also unintentionally neglect aspects of data, e.g., the Attribute Panel may not be able to fit all attributes of a dataset, causing the attributes that are outside the viewport to be potentially neglected during analysis. \app helps the user tackle both: by showing unintentionally uninteracted attributes and by allowing users to intentionally set custom target distributions for attributes.}

\subsection{Toward additional mitigation strategies}
Based on \app results, interaction traces help increase the user's awareness of their analysis practices, sometimes influencing them to interact and mitigate unconscious \remove{biases}\revise{biased analytic behaviors}. We believe this is a \emph{passive} mitigation strategy since the user has to inspect the difference between their \revise{analytic behavior} and the target distribution and devise an appropriate strategy, e.g., by applying a filter. \pa{01} and \pa{08} suggested we implement a more \emph{active} mitigation feature with \emph{``a button to automatically apply a reverse filter [instead of them having to manually apply it]''}, \emph{``especially for continuous attributes''}. For example, \pa{06} saw their interactions with different Genres (Concert, Documentary, and Western) and reflected \emph{``[they] should now interact with Drama since that is maximum and these are almost nil''}. They applied a filter to correct their \remove{unconscious bias}\revise{unintended underemphasis} but after a few interactions found themselves \revise{overemphasizing}\remove{biased} \emph{towards} Drama movies and reversed the filter. This act of balancing focus across all attributes can lead to frustration, sidelining the analysis task at hand. 
This is motivation to build mixed-initiative systems that assist the user in mitigating \remove{biases}\revise{biased analytic behaviors} either by acting on-demand (when the user clicks a ``Mitigate'' button for an attribute) or actively by automatically applying (or removing) a set of filters that negate the overemphasis (or underemphasis).

\subsection{Lessons Learned}
\remove{Grounded in the experiences designing, implementing, and evaluating \app, we share the following lessons learned.}

\bpstart{Encourage users to get lost in their analysis, but use awareness features to remind them.}
As described by the guidelines of ``fluid interaction'' \cite{elmqvist2011fluid}, users may become less aware of their own process while performing in-depth analysis. 
Achieving this level of usability and utility in visualization tools is desirable, but raises the need for awareness functionality such as that in \app. 
Awareness features can help remind users that alternatives should be considered. 

\bpstart{Awareness of one's own activity is helpful, guidance towards best ways to mitigate may be better.}
Participants saw utility in interaction traces toward increasing awareness of their analysis processes. But, what's next? While users may be aware of potential biased analytic behaviors, there may not be a clear path forward to correct those. Thus it can be fruitful to explore guidance to help users actively mitigate \remove{biases}\revise{biased analytic behaviors} (e.g., by recommending data, visualizations, or filters that may draw a user's attention to overemphasized or underemphasized parts of the data).

\bpstart{Different tasks call for different target distributions.}
``Biased analytic behaviors'' (deviations from a baseline) in context of a movies dataset can likely be chalked up to relatively harmless preferences.
However, given different analysis tasks, or different domain contexts (hiring, medical, etc), there may be a much more urgent need to ensure that the target baseline distribution is fit to the task. 
\app can provide the flexibility to specify custom target distributions accordingly.

\bpstart{Promote awareness while maintaining user agency and control.}
While we maintain that user agency and control should be ensured, providing people with awareness of their analysis behavior \remove{for the purpose of mitigating potentially detrimental biases} has merit. 
At times, these goals may be at odds. For instance, if active strategies are employed for systems to automatically apply bias-mitigating measures, then agency may be compromised.
These design decisions should be carefully considered when designing visualizations.




\section{Limitations}
\label{section:limitations}
\app currently supports only a small set of visualization types; however, we chose them to test across different aggregation types. 
Also, \revise{analytic behavior} is modeled only from interactions, which may not be a complete proxy for attention; in the future, one may consider user gaze or other sources to more accurately approximate it. Lastly, \app models \revise{analytic behavior} by equally weighting the interactions. As \pa{06} suggested to \emph{``remove older interactions, say only keep the most recent 100 or 200 of them [as] people lose attention [over time]''}, we may consider interaction recency in the future.



Finally, interactions with aggregate visualizations (e.g., hovering on a bar in a bar chart) are currently considered as N equally weighted interactions of magnitude 1/N where N = number of data points belonging to that element.
This has variable impact on the metrics due to different statistical tests used to compute the \remove{focus}\revise{analytic behavior model} (AD~\cite{wall2017warning}) depending on the attribute type (e.g., $\chi^2$ test for categorical attributes, Kolmogorov-Smirnov test for numerical distributions).
Future work can explore alternative computations for \remove{focus metrics}\revise{models of analytic behavior} that may reflect a user's attention and intentions more precisely.

\section{Conclusion}
\label{section:conclusion}
\app is a visual data analysis tool that models \revise{analytic behavior} of users from their interaction history and provides real-time feedback for awareness and self-reflection of\revise{, e.g., overemphasis (or underemphasis) on aspects of data}\remove{potential implicit biases}. Our evaluation found that \app increases users' awareness of their analysis behaviors in real-time, promoting reflection upon and acknowledgement of their intentions with the data.  
These results can have far-reaching implications toward mitigating \remove{bias}\revise{biased analytic behaviors} in decision making contexts, e.g., \revise{aid a hiring committee to meet their gender diversity targets}\remove{gender biases in hiring, racial biases in criminal investigations} and generally foster more transparent analysis processes.

\acknowledgments{
  This work was supported in part by the National Science Foundation grant IIS-1813281 and the Siemens FutureMaker Fellowship. We thank the reviewers for their constructive feedback during the review phase. We also thank the Georgia Tech Visualization Lab for their feedback.
}

\bibliographystyle{abbrv-doi}

\bibliography{template}
\end{document}